\documentclass{bioinfo}
\usepackage{amsmath}
\usepackage{lmodern}
\usepackage{float}
\copyrightyear{2018} \pubyear{2018}

\access{Advance Access Publication Date: Day Month Year}
\appnotes{Manuscript Category}

\begin{document}

\firstpage{1}

\subtitle{Subject Section}

\title[BRCA1 and BRCA2 SNP VUS  prediction]{Predicting clinical significance of BRCA1 and BRCA2 single nucleotide substitution variants with unknown clinical significance using probabilistic neural network and deep neural network-stacked autoencoder}

\author[Sample \textit{et~al}.]{Ehsan Rahmatizad KhajePasha\,$^{\text{\sfb 1}}$, Mahdi Bazarghan\,$^{\text{\sfb 1,}*}$, Hamidreza Kheiri Manjili\,$^{\text{\sfb 2}}$$^{\text{\sfb ,3}}$, Ramin Mohammadkhani\,$^{\text{\sfb 1}}$ and Ruhallah Amandi\,$^{\text{\sfb 1}}$}

\address{$^{\text{\sf 1}}$ Department of Physics, University of Zanjan, Zanjan, 45371-38791, I. R. Iran. and \\
$^{\text{\sf 2}}$Cancer Gene Therapy Research Center, Zanjan University of Medical Sciences, Zanjan, I. R. Iran. and \\
$^{\text{\sf 3}}$Zanjan Pharmaceutical Nanotechnology Research Center,
Zanjan University of Medical Sciences, Zanjan, I. R. Iran.
}

\corresp{$^\ast$Mahdi Bazarghan}

\history{Received on XXXXX; revised on XXXXX; accepted on XXXXX}

\editor{Associate Editor: XXXXXXX}

\abstract{\textbf{Motivation:}  Non-synonymous single nucleotide polymorphisms (nsSNPs) are single nucleotide substitution occurring in the coding
region of a gene and leads to a change in amino-acid sequence of protein. The studies have shown these variations may be associated with disease.
Thus, investigating the effects of nsSNPs on protein function will give a greater insight on how nsSNPs can lead into disease.  Breast cancer is the
most common cancer among women causing highest cancer death every year. BRCA1 and BRCA2 tumor suppressor genes are two main candidates of which, mutations
in them can increase the risk of developing breast cancer. For prediction and detection of the cancer one can use experimental or computational methods, but
the experimental method is very costly and time consuming in comparison with the computational method. The computer and computational methods have been used for more than 30 years.
Here we try to predict the clinical significance of BRCA1 and BRCA2 nsSNPs as well as the unknown clinical significances. Nearly 500 BRCA1 and BRCA2
nsSNPs with known clinical significances retrieved from NCBI database. Based on hydrophobicity or hydrophilicity and their role in proteins' second structure,
they are divided into 6 groups, each assigned with scores. The data are prepared in the acceptable form to the automated prediction mechanisms,
Probabilistic Neural Network (PNN) and Deep Neural Network-Stacked AutoEncoder (DNN).\\
\textbf{Results:} The preprocessed data is used for training and testing the PNN and DNN. With Jackknife cross validation we show that the prediction accuracy achieved
for BRCA1 and BRCA2 using PNN are 87.97\% and 82.17\% respectively, while 95.41\% and 92.80\% accuracies achieved using DNN. The total required processing time for the training and testing the PNN is 0.9 second and DNN requires about 7 hours of training and it can predict instantly. both methods show great improvement
in accuracy and speed compared to previous attempts. The promising results imply that the intelligent methods are suitable and applicable to such problems
without human interference, with very high prediction speed.\\
\textbf{Availability:} XXXXX \\
\textbf{Contact:} \href{bazargan@znu.ac.ir }{bazargan@znu.ac.ir }\\
\textbf{Supplementary information:} Supplementary data are available at \textit{Bioinformatics}
online.}
\maketitle
\section{Introduction}
Nowadays Artificial Intelligence (AI) plays an important role in science (e.g. chemistry, physics, medicine and etc.). While experimental methods are more reliable but they are also more time consuming and expensive than computational methods. Although computational methods may never get accurate enough to replace experimental methods, but they can help selecting and prioritizing to a small number of likely candidates from pools of available data. One of the computational methods used in medicine is Artificial Neural Networks (ANNs) which is a computer program and inspired from animal's nervous system. ANNs consist of simple processing units, nodes
(neurons), which aggregates and processes according to its internal activation function, then these units produce outputs. Every ANN consists of layers including an input layer, some intermediate layers and an output layer. The units (nodes) of each layer has connection with the nodes of its next layer, although more complex
patterns of connections are possible. ANNs can be divided into two main groups, supervised learning ANNs, and unsupervised learning ANNs.
Supervised learning ANNs need to be trained using examples of the problem under investigation. In the training process,
the weights relating each connection between units and parameters of the activation function in each unit of the network are
adjusted in a direction to reduce the output error. By training an ANN with the training set, which is a set of different, but
related input patterns, The ANN can relate input to output without using explicit algorithms for deciding the appropriate output.\\
ANNs have been used in cancer detection and diagnosis for more than 30 years 
(\citealp{Sim85},\citealp{Mac91},\citealp{Cic92}) but as noted in a study, the fundamental goals of cancer prediction and prognosis are different from the goals
of cancer detection/diagnosis. In cancer prediction one tries to (i) predict cancer susceptibility, (ii) predict cancer recurrence, (iii) predict cancer survivability. 
Prediction is more useful than detection because in prediction one can get prepared before the occurrence of cancer. Also in the study it has been noted that almost
all predictions use just four types of input data (i) genomic data, (ii) proteomic data, (iii) clinical data, (iv) combination of these data (\citealp{Cru06}).\\
A non-synonymous SNP (nsSNP) is a single nucleotide substitution occurring inside coding region of a gene, causing an amino acid substitution in the corresponding protein product.
These changes can lead to a structural or functional change in the protein product which may give a minor or major phenotypic change or may absolutely have no effect. For example,
a nsSNP in the hemoglobin beta gene (substitution of glutamic acid by Valine) is one cause for sickle cell anemia (\citealp{Wes01}), also diabetes has been correlated with a number of
nsSNPs (\citealp{Sha05}). A mutation can affect protein folding and stability, protein function, protein-protein interaction, protein expression 
and sub-cellular localization (\citealp{Rev11})
and they can be divided into two categories (i) apparently random (Sporadic) mutations followed by somatic selection (somatic mutations), (ii)
pre-existing mutations in the germline (germline mutations). Mutations can occur due to (i) gain of function mutations that changes the normal
gene into an oncogen. (ii) Loss of function mutations that inactivates tumor suppressor genes. (iii) drug resistance mutations that overcomes
the inhibitory effect of a drug on the targeted protein. Nonsynonymous single nucleotide polymorphisms (nsSNPs) are prevalent in genomes and
are closely associated with inherited diseases. To facilitate identifying disease-associated nsSNPs from a large number of neutral nsSNPs, it
is important to develop computational tools to predict the nsSNP's phenotypic effect (disease-associated versus neutral).\\
Breast cancer is the
most common cancer among women causing highest cancer death every year(\citealp{xei2016},\citealp{xei2017},\citealp{xei2018}). The major susceptibility genes, BRCA1 and BRCA2, acting as a tumor suppressor have been previously identified (\citealp{Ant00},\citealp{Kur10}). Pathogenic mutations in
these genes increase the inherited predisposition to breast cancer. Evidence are suggesting that genetic variants may alter the breast cancer
risk for those with BRCA1 or BRCA2 mutations. There is a study reporting a relationship between a woman BRCA2 SNPs profile and the age she
develops breast cancer (\citealp{Joh13}). Another study on individuals carrying inactivating germline mutations in BRCA1 shows that they have an increased
risk of developing cancer (\citealp{Shat97}). So it's essential to identify those at risk. The risk estimates of developing breast cancer in a woman who
carries a BRCA1 or BRCA2 mutations is from kindred with multiple cases of breast or ovarian cancer, or both, range from 76 to 87\% (\citealp{Cru06}).\\
Variants with Unknown Clinical Significance (VUS) in BRCA1 or BRCA2 genes cause major problems because physicians do not know whether the VUS
is related to developing BC or is neutral with respect to BC risk. Thus carriers of VUSs cannot benefit from risk assessment, prevention and
therapeutic measures that are available to carriers of known significance mutations. The recent increase in the available nsSNP data, determining
the clinical significance of VUS in BRCA1 and BRCA2 has become an important clinical issue and doing it automatically using the ANN is valuable. Here we use the PNN, which is a supervised learning ANN and benefits from speed and easy result interpretation. The network
is trained with nsSNP data of BRCA1 and BRCA2 to predict the clinical significance of VUSs. We retrieve data from NCBI 
\footnote{\vspace{5pt}ncbi.nlm.nih.gov/snp} 
with 449 nsSNP data of
homo sapiens BRCA1 and 460 nsSNP data of homo sapiens BRCA2 and then preprocess them. We train PNN and use different methods of validation (e.g.
jackknife and cross-validation) and achieve different accuracies and use best data model to train and test the DNN. We show that, given enough data, both PNN and DNN can outperform other ANN algorithms
at accuracy and speed. Larger the training samples, more will be the accuracy.
\enlargethispage{12pt}

\begin{methods}
\section{Methods}
\subsection{Probbablistic neural network (PNN)}
\begin{figure}[t]
	\centerline{\includegraphics[width=160pt,height=95pt]{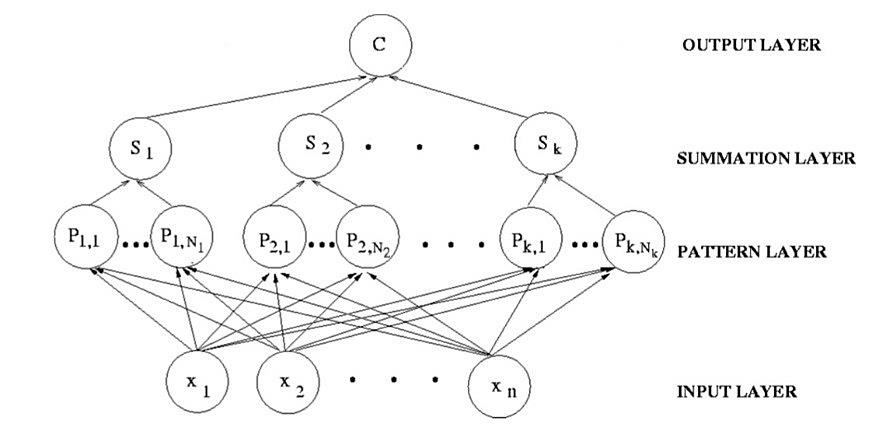}}
	\vspace*{-5pt}
	\caption{The architecture of a typical PNN.}\label{fig:01}
\end{figure}
Strategies used to classify patterns in a way that they minimize the expected risk are called "Bayes strategies".
PNN introduced by Specht is a result of the theory of statistical pattern classification. In the fifties and sixties parametric
methods used to solve statistical pattern classification problems but within last twenty years
these methods are replaced by the non-parametric approach (\citealp{Rut04}).
In the non-parametric method, it is assumed that a functional form of probability densities is unknown. Pattern classification
procedures derived from non-parametric estimates are convergent - when the length of learning sequence increases - to Bayes' rules.
PNNs are applied in many interesting fields and they implement non-parametric estimations
techniques in parallel fashion to benefit from fast training and convergence to Bayes optimal decision surface (\citealp{Rut04}).
The architecture of a typical PNN is as shown in fig.\ref{fig:01}.\\
The input layer without any computation distributes the input neurons in the pattern layer. Neuron $x_{ij}$ after receiving a pattern x computes its output.
\begin{equation}
\phi_{ij}(x)=\frac{1}{(2\pi)^{d/2}\sigma^d}exp[-\frac{(x-x_{ij})^T(x-x_{ij})}{2\sigma^2}]\label{eq:01}\vspace*{-10pt}
\end{equation}\\
where d denotes the dimension of the pattern vector x, $\sigma$ is the smoothing parameter and $x_{ij}$ is the neuron vector and by summarizing and averaging the
output of all neurons belonging to the same class, the summation layer neurons
compute the maximum likelihood of pattern x being classified into $C_i$.
\begin{equation}
P_{i}(x)=\frac{1}{(2\pi)^{d/2}\sigma^d}\frac{1}{N_i}\sum_{j=1}^{N}exp[-\frac{(x-x_{ij})^T(x-x_{ij})}{2\sigma^2}]\label{eq:02}\vspace*{-10pt}
\end{equation}\\
$N_i$ denotes the total number of sample in class $C_i$. The decision layer unit will classify the pattern x in accordance with the Bayes'
decision rule based on the output of all the summation layer neurons if the a priori probabilities and losses associated with making an incorrect
decision for each class are the same (\citealp{Rut04}).
\begin{equation}
\hat{C}(x)=arg\hspace{2pt}max\{p_i(x)\} ; i = 1 , 2 , ... , m\label{eq:03}\vspace*{-10pt}
\end{equation}\\
Where $\hat{C}(x)$ denotes the estimated class of the pattern x and m is the total number of classes in the training samples.
\subsection{Deep Neural Network-Stacked AutoEncoder}
\begin{figure}[t]
	\centerline{\includegraphics[width=160pt,height=150pt]{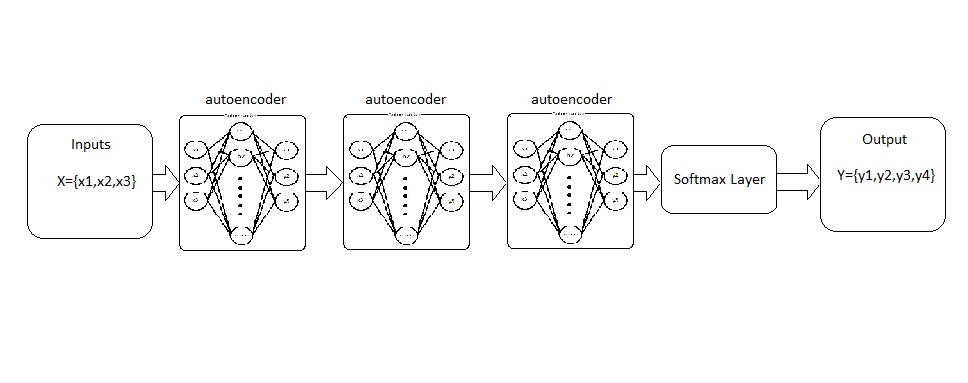}}\vspace*{-50pt}
	\caption{Deep Learning Structure used in this work.}\label{fig:03}
\end{figure}
Todays modern technology influenced by many aspects of machine-learning, from video content analysis to large scale image processing(\citealp{deep1}), 
from self-driving car(\citealp{deep2}),to recommended system(\citealp{deep3}).
From the beginning the research intention of pattern recognition was to replace human engineered features by multilayer neural network,
no suitable algorithm provided for this goal since 1980s decade. The backpropagation used to compute the gradient of an objective function by optimizing weights(\citealp{deep4}).\\
Learning an undirected graphical model called a Gaussian binary restricted Boltzmann machine (RBM) that has one visible layer of
linear variables with Gaussian noise and one hidden layer. There is full connectivity between layers. The connection weights
(and biases) can be learned efficiently using the contrastive divergence approximation to the log likelihood gradient(\citealp{deep5}).\\
Auto-encoders are simple building-block for learning, each of these blocks transform input data into outputs to create more
efficient form of data. One can build up most powerful machine learning algorithm by combining these simple parts. Auto-encoders were first introduced by Hinton for
unsupervised learning of back-propagation algorithm(\citealp{deep6}). After many years auto-encoders take inner place in deep architectures, where stacked auto-encoders combined with 
supervised top-layer. These deep architectures show state-of-art results for many challenging problems(\citealp{deep8},\citealp{deep9}).
The Encoder building block is given in figure \ref{fig:03}
Auto-encoders, mainly, are of two types; linear auto-encoder and non-linear auto-encoder. Linear auto-encoder is equivalent to Principle Component Analysis (PCA),
also by using nonlinear transfer function one can discover nonlinear patterns of data. Boolean and Boltzmann learning machines are famous non-linear models.
De-noising auto-encoder is other kind of auto-encoder that reconstructs corrupted patterns(\citealp{deep9}).\\
Auto-encoder can be defined by set of parameters (n, p, N, F, G, A, B, X, Y, D) 
n and p are the positive integers, which show the number of units for input/output and hidden layer. N is the number of training samples and
A is the class transfer functions that relates G in hidden units to F in output layer.
B is the transfer functions that relates input units in set F to hidden units.
\begin{math}X\in R^1\end{math} is an input vector for auto-encoder; auto-encoder converts the vector X into another similar vector as follows:
\begin{equation}
	z=h(wx+b)\label{eq:04}\vspace*{-10pt}
\end{equation}
h is the transfer function for each layer of auto-encoder, w is a weight matrix and b is a bias vector, the decoder reverses this process and
converts back the z to x vector(\citealp{deep10}). Y is the target vector. D measures the similarity over input layer units.\\
If we define $A_1$ as a subset of A and $B_1$ as a subset of B then for any input we would like to find the parameters to transform the input units in
$A_1$ to output units in $B_1$ in which we want to minimize the solution for: 
\begin{equation}
\begin{split}
min E(A_1,B_1)=& min(A_1,B_1)\sum_{i=1}^{N} E(A_1,B_1)= \\
             & \sum_{i=1}^{N} D(f_{A_1,B_1}(x_{target}),y_{target})
\end{split}
\label{eq:05}\vspace*{-10pt}
\end{equation}
Where $f_{AB}$ is the function that related A to B, D is similarity measure and N is the number of training data (\ref{eq:05}).\\
We can add regularization to the cost function(equation1) and create sparsity for auto-encoder. To compute cost functions for encoder,
usually $L_2$ and sparsity regularization terms  combine with mean square error term.\\
The hidden layers in auto-encoders are of two types named as compressed and sparse. In compressed type, number of units in hidden layer
is less than input layer, so network tries to combine features and reproduce new compressed features. But in sparse auto-encoder, features are expanded (\citealp{deep11}).
The network structure which is used in this study is given in figure \ref{fig:03}.
Here we defined simple stacked auto encoder as deep network to analyze our dataset. We used three simple auto-encoders with hidden node of size 300 in each one,
these layers combined with one softmax layer to create deep structure.\\
\subsection{Data}
Here, we used ncbi.nlm.gov/snp among all the nsSNP databases because it is complete and also details are provided with each data. for now, we only used nsSNPs
and didn't include indel variations. We used our program to pre-process data which translates DNA code to protein code (transcript ID for BRCA1: NM 007294,
transcript ID for BRCA2: NM 000059.3). The final step was to make a good coding scheme for amino-acids (AAs) so that PNN can achieve high accuracy at predicting
the clinical significance. We used hydropathy and propensity indices for scoring AAs and divided them into 6 categories based on if an AA is hydrophobic or
hydrophilic and if it favors alpha-helices or turns or beta sheets in its 2nd structure.
\subsubsection{BRCA1}
We retrieved 5591 data from the database and used our program to apply variants, after applying nsSNP data our program showed 1871 of
them are the exonic variants of which 449 were variants with known clinical significance. So we used these 449 data to train and test the PNN and DNN. We tried out 5 different methods to prepare
our data and each time improved it. Five different types of prepration used for data are as below:\\
1 - After application of SNPs, DNA sequences are converted to number string using this scheme : a  = 1 , c = 2 , g = 3 , d = 4.\\
2 - DNA sequences are converted to number string using this scheme : All nodes are zero except for the node which is different from main sequence node.\\
3 - DNA sequences are translated to AA sequences according to their transcription ID and then AA sequences are converted to number strings according to the scheme used to classify AAs.\\
4 - AA sequences are converted to number strings where all digits are zero except changed ones. \\
5 - This model is the most important because it gave promising results of accuracy and speed due to its short length and amount of data fitted into its nodes. In this model we used just 3 nodes to represent every SNP, first node dedicated to the location of AA substitution caused by SNP while second and third nodes are 
for old AA and new AA. Every data in this model contains 5 different information involving, AA substitution location , old and new AA's hydrophilicity or hydrophobicity and
their favorite structure.
\subsubsection{BRCA2}
The same procedure was applied to BRCA2 data. As a raw data, we had 7227 BRCA2 nsSNP, then after processing them with our
program we had 2972 coding nsSNP data. The program showed that 460 of them are variant with known clinical significance.

\section{Results}
Training and testing with 5th dataset resulted in the highest accuracy that we could get out of PNN and DNN. The reason why this dataset is the most suitable one is simple.
Because it has the shortest possible length for representing AA substitution and provides enough information for AI so that it can classify AA substitutions with sufficient accuracy.\\
Table \ref{Tab:01} shows PNN accuracy at predicting AA substitution and table \ref{Tab:02} shows same results obtained using DNN.
\begin{table}[H]
	\processtable{Accuracies achived training PNN using different datasets\label{Tab:00}} {\begin{tabular}{@{}lll@{}}
			\toprule dataset number$^{*}$ & BRCA1(\%) & BRCA2(\%)\\
			\midrule
			1 & 48.2 & 40.4\\
			2 & 50.4 & 45.7\\
			3 & 55.6 & 48.8\\
			4 & 58.7 & 53.2\\\botrule
	\end{tabular}}{*prediction accuracies using 5th dataset shows significant increase in accuracies which are given at table \ref{Tab:01} and table \ref{Tab:02}.}
\end{table}
\vspace{-20pt} 
\begin{table}[H]
\processtable{PNN accuracy at predicting BRCA1 and BRCA2 SNP's clinical significance\label{Tab:01}} {\begin{tabular}{@{}lll@{}}
\toprule Validation method & BRCA1(\%) & BRCA2(\%)\\
\midrule
5-fold cross validation & 78.40 & 79.13\\
10-fold cross validation & 80.40 & 80.00\\
20-fold cross validation & 85.45 & 80.22\\
Jackknife & 87.97 & 82.17\\\botrule
\end{tabular}}{}
\end{table}
\vspace{-19pt}
\begin{table}[H]
\processtable{DNN accuracy at predicting BRCA1 and BRCA2 SNP's clinical significance\label{Tab:02}} {\begin{tabular}{@{}lll@{}}
\toprule Validation method & BRCA1(\%) & BRCA2(\%)\\
\midrule
5-fold cross validation & 86.74 & 79.95\\
10-fold cross validation & 92.04 & 81.14\\
20-fold cross validation & 93.30 & 82.86\\
Jackknife & 95.41 & 92.80\\\botrule
\end{tabular}}{}
\end{table}
\vspace*{-45pt}
\begin{table}[H]
	\processtable{PNN and Deeplearning evalution benchmarks for BRCA1\label{Tab:03}} {\begin{tabular}{@{}lll@{}}
			\toprule Benchmark \hspace{30pt} & PNN\hspace{20pt} & DNN\\
			\midrule
			Accuracy & 87.97\% & 95.41\%\\
			Sensitivity & 93.96\% & 79.73\%\\
			Specificity & 62.35\% & 93.87\%\\
			F1-score & 0.9268 & 0.8059\\
			MCC & 0.5914 & 0.7454\\\botrule
	\end{tabular}}{MCC : Matthews correlation coefficient}
\end{table}
\vspace*{-20pt}
\begin{table}[H]
	\processtable{PNN and Deeplearning evalution benchmarks for BRCA2\label{Tab:04}} {\begin{tabular}{@{}lll@{}}
			\toprule Benchmark \hspace{30pt} & PNN\hspace{20pt} & DNN\\
			\midrule
			Accuracy & 82.17\% & 92.80\%\\
			Sensitivity & 87.72\% & 64.01\%\\
			Specificity & 50.72\% & 90.12\%\\
			F1-score & 0.8932 & 0.6723\\
			MCC & 0.3570 & 0.5807\\\botrule
	\end{tabular}}{}
\end{table}
\vspace*{-35pt}
\end{methods}
\section{Discussion}
In this article, we have shown that choosing the right data (right tool for the right job) will change the results. As shown in the table \ref{Tab:00} when we had chosen DNA sequence instead of protein sequence our result would not be more than 50.4\% and 45.7\% for BRCA1 and BRCA2 respectively.\\
As reported in (\citealp{disc01}) several studies have considered how benign and pathogenic nsSNPs may be distinguished using only sequence and structural aspects of the proteins in which they occur, e.g. Wang and Moult in (\citealp{disc02}) used protein hydrophobic core disruption to determine the proteins structural stability indirectly. Here we have used hydrophobicity and hydrophilicity alongside favorite second structure of the previous AA and new AA to score the AAs substitution. Our results suggest that this information will lead to the increase of the accuracy of prediction making machine-learning methods a useful tool for SNP prediction problems in a way that our method can be practical to real-world applications. A clear limitation of our study was the inability to use indel SNPs due to assumptions made in the definition of the model.\\
Prediction accuracy difference for BRCA1 and BRCA2 is because BRCA1 AA sequence length is 1863 AA while BRCA2 AA sequence consists of 3418 AAs. Since we have almost the same number
of AA substitution with known clinical significance for both BRCA1 and BRCA2 , fraction of our knowledge which we use to train AI for BRCA1 is almost twice the knowledge that we have
about BRCA2.\\
In order to evaluate the behavior of NNs, we have used five well-known measures: accuracy, sensitivity, specificity, F1-score, Mathews Correlation Coefficient (MCC). In general, sensitivity indicates how well the NN can predict the actual positive (e.g. a pathogenic sample as a pathogenic sample) and specificity indicates how well the NN can identify the actual negative (e.g. a benign sample as a benign sample). The F1-score is the harmonic average of the precision and sensitivity. Best F1-score is 1 while the worst F1-score is 0. MCC is a measure to indicate the quality of learning where value of +1 shows perfect prediction and value of 0 shows prediction quality no better than r+1andom prediction and -1 shows a prediction in total disagreement between prediction and observation. Values of each measure for each gene and NN used are shown in tables \ref{Tab:03} and \ref{Tab:04}.\\
As the result, it is shown that DNN can predict AA substitutions with more accuracy due to its modern structure and its efficiency.\\
According to (\citealp{Cru06}) the lack of attention to the data validation is one of the major problems in this field. As reported in the article, 5-fold or 10-fold cross validation is sufficient for most learning algorithms to be validated, however, we used a more aggressive method to validate the learning quality of the NNs which is jackknife cross-validation method in this process we have used all but one sample iteratively to train the NNs. Also, a common problem in this field is the imbalance problem, in which the data set is dominated by a major class, so the predictions have a bias toward that class. To check if our method is affected by this problem or not, we performed the learning and testing procedure using the same number of samples from each class. The obtained result showed a minor (2-5\%) change in the accuracy of the final result that reported in the table \ref{Tab:02}.\\
In addition to the supervised learning algorithms reported here, we have also use Self-Organizing Maps (SOM) to check whether these algorithms can or can't distinguish between different classes. Resulted Maps suggested the inability of these algorithms in the classification of nsSNPs.\\
%
%
\section{Conclusion}
In this article, we have used PNN and DNN to predict the clinical significance of nsSNPs with unknown clinical significance. Our data were obtained from NCBI database and then we have used DNA sequence and Protein sequence and 2nd structural information to prepare our training and testing data set. Among different datasets, protein dataset with a novel model of showing nsSNP position and substitution showed best results, then we used n-fold cross validations to validate our results. Also, F1-score and MCC scores are reported to show the quality of learning and prediction of NNs used.\vspace*{-10pt}
\section*{Acknowledgements}

We thank Dr Emarn Heshmati and Dr S. Shahriar Arab for their guides at the beginning of this research.   \vspace*{-12pt}



\bibliographystyle{natbib}
\bibliographystyle{bioinformatics}
%
%

\end{document}